# Data-Driven Real-Time Power Dispatch for Maximizing Variable Renewable Generation

Zhigang Li[a,*], Feng Qiu[b], Jianhui Wang[b]
[a] *Department of Electrical Engineering, Tsinghua University, 100084, Beijing, China*
[b] *Energy Systems Division, Argonne National Laboratory, 60439, Argonne, USA*

***Abstract*—Traditional power dispatch methods have difficulties in accommodating large-scale variable renewable generation (VRG) and have resulted in unnecessary VRG spillage in the practical industry. The recent dispatchable-interval-based methods have the potential to reduce VRG curtailment, but the dispatchable intervals are not allocated effectively due to the lack of exploiting historical dispatch records of VRG units. To bridge this gap, this paper proposes a novel data-driven real-time dispatch approach to maximize VRG utilization by using do-not-exceed (DNE) limits. This approach defines the maximum generation output ranges that the system can accommodate without compromising reliability. The DNE limits of VRG units and operating base points of conventional units are co-optimized by hybrid stochastic and robust optimization, and the decision models are formulated as mixed-integer linear programs by the sample average approximation technique exploiting historical VRG data. A strategy for selecting historical data samples is also proposed to capture the VRG uncertainty more accurately under variant prediction output levels. Computational experiments show the effectiveness of the proposed methods.**

## Nomenclature

*Sets*

| | |
|---|---|
| $\mathcal{D}^{\text{hist}}$ | Index set of historical data of forecast errors of variable renewable generation (VRG). |
| $\mathcal{G}^{\text{CC}}$ | Index set of corrective control units. |
| $\mathcal{G}^{\text{NCC}}$ | Index set of non-corrective control units. |
| $\mathcal{G}^{\text{VRG}}$ | Index set of renewable energy generation (VRG) units. |
| $\mathcal{G}_n^{(\cdot)}$ | Index set of $(\cdot)$ -type units connected to bus *n*. |
| $\mathcal{I}^{\text{bus}}$ | Index set of buses. |
| $\mathcal{I}^{\text{line}}$ | Index set of transmission lines. |
| $\mathcal{S}^{\text{DNE}}$ | Index set of data samples for the do-not-exceed (DNE) limit determination problem. |
| $\mathcal{S}^{\text{OBP}}$ | Index set of data samples for the operating base point (OBP) optimization problem. |

*Parameters and Functions*

| | |
|---|---|
| $C_i(\cdot)$ | Generation cost of generation unit *i*. |
| $D_n$ | Load demand at bus *n*. |
| $e_j^{(k)}$ | *k*-th sample of forecast error of VRG unit *j*. |
| $F_l^{\max}$ | Transmission capacity of line *l*. |
| $LMP_j$ | Locational marginal price at the bus connected to VRG unit *j*. |
| $N^{\text{s}}$ | Number of selected samples. |
| $P_i^{\min}$ | Minimum output of generation unit *i*. |
| $P_i^{\max}$ | Maximum output of generation unit *i*. |
| $p_i^0$ | Generation output of unit *i* in the current period. |
| $p_i^*$ | Base point of generation unit *i*. |
| $SF_{l,n}$ | Generation shift factor of line *l* to bus *n*. |
| $W_j^{\text{f}}$ | Forecast available output of VRG unit *j*. |

* Corresponding author. Present address: Rm. 3-120, West Main Building, Tsinghua University, 100084, Beijing, China. Tel: +86 138 1049 1952. E-mail address: lizg163@126.com.

| | |
|---|---|
| $W_j^{\max}$ | Generation capacity of VRG unit $j$. |
| $\varepsilon$ | Convergence tolerance. |
| $\Delta_i$ | Maximum corrective output adjustment of unit $i$. |

*Decision Variables*

| | |
|---|---|
| $l_j$ | Lower DNE limit of VRG unit $j$. |
| $p_i^{\mathrm{B}}$ | OBP of generation output of conventional unit $i$. |
| $p_i^{\mathrm{C}}$ | Corrective dispatch output of conventional unit $i$. |
| $\boldsymbol{s}$ | Slack variable. |
| $u_j$ | Upper DNE limit of VRG unit $j$. |
| $w_j$ | Generation output of VRG unit $j$. |
| $z^{(k)}$ | Binary indicator variable that is equal to 0 if the $k$-th forecast error sample lies within the DNE limits and equal to 1 otherwise. |
| $\alpha_j^{(m)}$ | Auxiliary binary indicator variable. |
| $\beta_j^{(m)}$ | Auxiliary binary indicator variable. |

*Random Variables*

| | |
|---|---|
| $\tilde{v}_j$ | Coefficient that denotes the actual generation output of VRG unit $j$ as a convex combination of the DNE limits. |
| $\tilde{w}_j$ | Actual available output of VRG unit $j$. |

*Acronyms*

| | |
|---|---|
| CCU | Corrective control unit. |
| DNE | Do-not-exceed. |
| ED | Economic dispatch. |
| ERCOT | Electricity Reliability Council of Texas. |
| IO | Interval optimization. |
| LMP | Locational marginal price. |
| MILP | Mixed-integer linear program. |
| MISO | Midcontinent Independent System Operator. |
| NCCU | Non-corrective control unit. |
| OBP | Operating base point. |
| RO | Robust optimization. |
| SAA | Sample average approximation. |
| UC | Unit commitment. |
| VRG | Variable renewable generation. |

## 1. Introduction

Large-scale variable renewable generation (VRG) is being integrated into the power grids for its economical and environmental benefits. In China, more than 90 GW of wind farms have been built and the wind power penetration rate is expected to reach 11% by 2020 [1]. In the United States, wind energy is expected to supply 20% of the electricity generation by 2030 [2]. Meanwhile, VRG brings significant challenges to power system operations owing to its inherent uncertainty and variability. Since it is difficult to forecast the availability of VRG accurately in advance, the traditional dispatch paradigm for conventional power sources would incur excessive VRG curtailment because it ignores potential transmission congestion and insufficient regulation capability of conventional units caused by VRG volatility. The curtailment levels in major U.S. regions have been up to 4% in the past years [2].

Recently, the power industry has launched initiatives to reduce wind curtailments by improving wind dispatchability. For example, the Electric Reliability Council of Texas (ERCOT) has redesigned market pricing and used faster generation schedules [3]. The Midcontinent Independent System Operator (MISO) implemented the Dispatchable Intermittent Resource protocol to perform an automatic process for wind curtailment [4].

Methodologies to promote VRG in power dispatch have also drawn much attention in the recent literature. To explore the impact of wind power uncertainty on power system scheduling, wind power uncertainty is described using scenario-based stochastic approach for unit commitment (UC) and economic dispatch (ED) in [5]. The multiple scenario approximation method is also employed to hedge the sub-hourly variability of renewable energy for stochastic UC problem in [6]. The accuracy of scenario-based stochastic optimization relies on the presumption of probabilistic distribution of random variables, which is difficult to describe accurately in practice. Alternatively, interval optimization (IO) method is used in [7] for operating strategy decision with wind power uncertainty, where interval numbers are employed to represent the variability of wind power. The IO method uses upper and lower bounds to describe the uncertain parameters, and derives optimistic and pessimistic solutions for satisfying the constraints under uncertainties. Wind power correlation is not effectively captured using such technique. Robust optimization (RO) is another option to formulate decision-making problems considering parameter uncertainty. In contrast to IO, the RO approach seeks the optimal solution that ensures the satisfaction of all constraints under any realization of uncertain parameters within the uncertainty sets. Given the possible realizations of uncertain parameters, the RO method yields an optimal solution that is robust to randomness, rather than ranges of solutions. In [8], two-stage RO is employed to model building energy scheduling considering chillers and ice thermal energy storage, and this method is shown to outperform the deterministic method. Reference [9] proposes an RO-based wind power dispatch framework for look-ahead dispatch. In this framework, wind farms are scheduled by using allowable power generation intervals, within which the system security can be guaranteed, as their dispatch guidance. In [10], this framework is extended to real-time power dispatch while the affine control rules of automatic generation control units are addressed. Similar to the allowable power generation intervals in [9] and [10], the do-not-exceed (DNE) limit is introduced in [11] to describe the maximum ranges of VRG that can be accommodated by the system without causing reliability issues, and a two-step dispatch framework using the DNE limits is also proposed. At the first stage of this dispatch framework, a conventional ED problem is solved to procure the operating base points (OBPs) as dispatch targets for conventional units at the next period. At the second stage, DNE limits of VRG units are calculated based on fixed OBPs and are used as dispatch signals for VRG units at the next period.

Power dispatch frameworks based on dispatchable intervals for VRG in [9]-[11] have been reported to outperform the conventional ED approach in terms of wind integration improvement with reliability guarantee, but several issues prevent these approaches from being further improved. First, the range allocation among different VRG units relies on the choice of weights, which might be misleading due the following reason. The weight factors can be predefined according to the operators' preference [9]-[10] or based on the locational marginal prices (LMPs) [11]. However, these strategies may not yield weight coefficients that reflect the actual demand of VRG units for flexibility. For example, in the LMP-weighted range calculation, a high LMP at a bus could possibly indicate congestion at that bus; thus, allocating a wider DNE range for VRG units at that bus would not be very helpful. The DNE limits are likely to be misled by the inappropriate choice of weights, and the VRG output could consequently be curtailed unnecessarily. Second, in [11], the DNE limits are calculated with fixed OBPs of conventional units. Since the regulation capability of conventional units is affected by the OBPs, fixing the OBPs might unnecessarily restrict the width of DNE ranges, thus providing an inefficient solution to VRG accommodation.

To address these issues, a real-time power dispatch methodology is developed to improve the utilization of VRG in this paper. Unlike the previous method [11] that calculates DNE limits with fixed OBPs, our approach co-optimizes the OBPs of conventional units and DNE limits of VRG units simultaneously to maximize the probability that the actual available VRG output is covered by the DNE limits. Without introducing any weight factors or making any assumptions about the probability distributions of VRG, a data-driven method is employed to model the DNE limit decision problem, exploiting the historical data of VRG output to capture its stochastic features.

The contributions of this paper is three-fold: (1) A novel real-time power dispatch framework for VRG is developed. (2) A data-driven approximation model is formulated to determine the DNE limits of VRG units without predefining weights. (3) A data selection strategy is proposed to obtain the most relevant samples for DNE limit calculation.

The remainder of this paper is organized as follows. In Section 2, a two-stage real-time power dispatch framework based on DNE ranges is presented. In Section 3, optimization models for DNE limit and OBP determination are formulated using data-driven approximation and a strong extended reformulation techniques. Solution algorithms are also discussed in this section. Section 4 describes a strategy to select data for computation. Numerical tests are conducted and simulation results are discussed in Section 5. Finally, conclusions are given in Section 6.

## 2. Real-Time Power Dispatch Framework for VRG

The concept of DNE limits was proposed to determine the ranges of VRG output within which system reliability can be maintained with corrective actions [11]. "DNE limit" refers to the maximum output ranges of VRG units that can be accommodated by the system without causing reliability issues. DNE limits for the next period are calculated and sent to the VRG units as dispatch guidance. An improved real-time power dispatch framework using the DNE limits is introduced in this section.

The real-time power dispatch procedure with VRG is performed in a receding horizon to provide the dispatch guidance for all controllable generation units at the next dispatch period. According to their different control modes, generation units are sorted into three types: corrective control units (CCUs), non-corrective control units (NCCUs), and VRG units. CCUs are able to take corrective control actions after the availability of VRG is realized, and NCCUs follow the OBPs. VRG units are subject to taking proactive measures to keep their output within the DNE limits.

Each run of the real-time dispatch procedure consists of three steps, as shown in Figure 1.

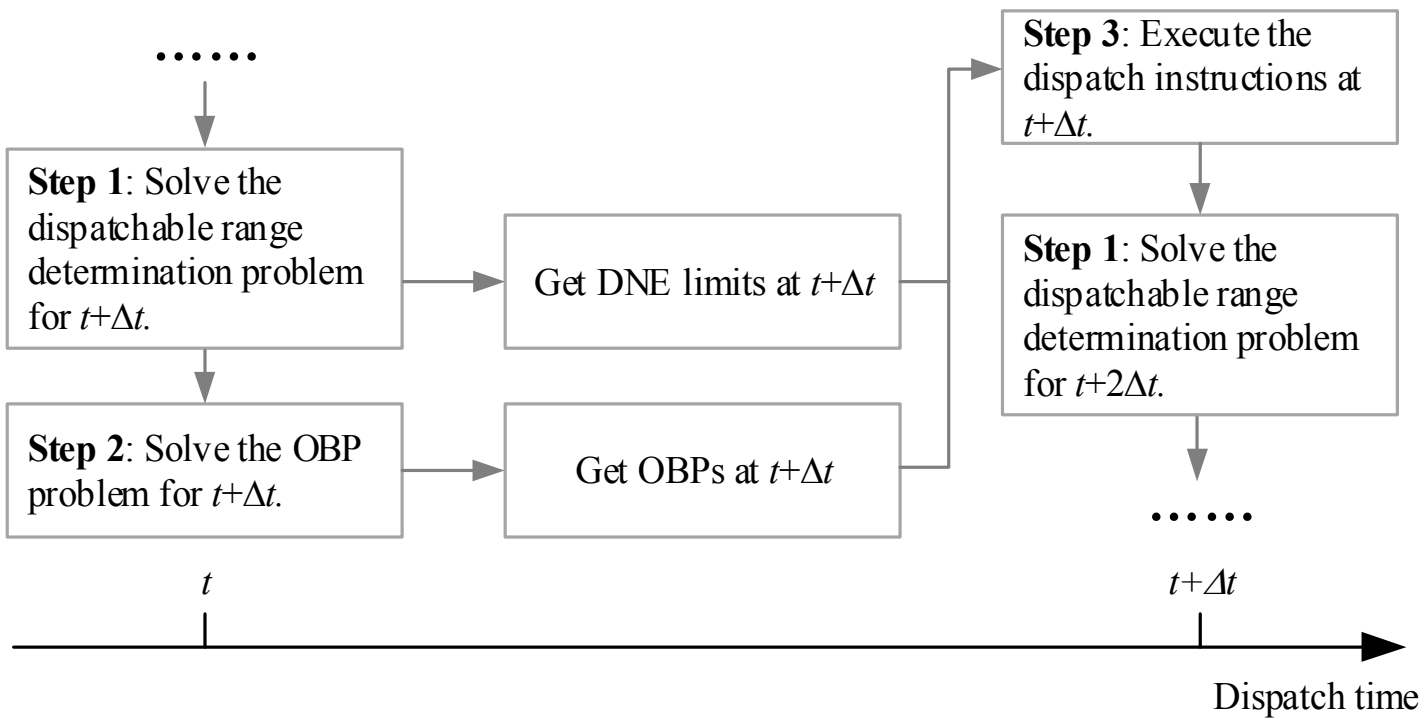

Figure 1. Real-time power dispatch framework for VRG using DNE limits.

**Step 1**: The dispatchable range determination problem is solved to obtain the DNE limits of VRG units in the upcoming period. The formulation and solution of the dispatchable range determination problem are discussed in detail in Section 3.

**Step 2**: The OBP optimization problem with fixed DNE limits is solved to get the OBPs of conventional units in the upcoming period. The formulation and solution of the OBP problem are also be discussed in detail in Section 3.

**Step 3**: The dispatch decisions are executed in the next period. CCUs provide corrective actions based on their OBPs, NCCUs follow their OBPs, and VRG units take proactive actions to keep their output within the DNE limits.

## 3. Dispatchable Range Determination and Calculation of OBPs

This section focuses on determining the DNE limits in Step 1 and the OBPs in Step 2 by using a data-driven approach. The dispatchable range determination and OBP optimization problems are first formulated in Section 3.1, the formulations of which are approximated in Section 3.2 by mixed-integer linear program (MILP) models using the sample average approximation (SAA) technique based on historical VRG forecast data. To make the MILP models more compact and tighter, a strong extended formulation technique is used to further reformulate the problems in Section 3.3. The solution algorithms for the final models are presented in Section 3.4.

### *3.1 Problem Formulation*

*1) Dispatchable range determination problem.* This problem seeks the ranges of VRG outputs that cover the most possible scenarios, within which system reliability can be maintained or restored by using corrective actions. The problem is formulated as follows:

**(DNE-1)**

$$\max_{\boldsymbol{l},\boldsymbol{u},\boldsymbol{w},\boldsymbol{p}^{\mathrm{B}}} \mathbb{P}\{\boldsymbol{l} \le \tilde{\boldsymbol{w}} \le \boldsymbol{u}\} \tag{1}$$

$$\text{subject to:} \quad \sum_{i\in\mathcal{G}^{\mathrm{CC}}\cup\mathcal{G}^{\mathrm{NCC}}} p_i^{\mathrm{B}} + \sum_{j\in\mathcal{G}^{\mathrm{VRG}}} w_j = \sum_{n\in\mathcal{I}^{\mathrm{bus}}} D_n\,, \tag{2}$$

$$\left|\sum_{n\in\mathcal{I}^{\mathrm{bus}}} SF_{l,n}\cdot\left(\sum_{i\in\mathcal{G}_n^{\mathrm{CC}}\cup\mathcal{G}_n^{\mathrm{NCC}}} p_i^{\mathrm{B}} + \sum_{j\in\mathcal{G}_n^{\mathrm{VRG}}} w_j - D_n\right)\right| \le F_l^{\max},\ \forall l\in\mathcal{I}^{\mathrm{line}}\,, \tag{3}$$

$$-RAMP_i \le p_i^{\mathrm{B}} - p_i^0 \le RAMP_i,\ \forall i\in\mathcal{G}^{\mathrm{CC}}\cup\mathcal{G}^{\mathrm{NCC}}\,, \tag{4}$$

$$P_i^{\min} \le p_i^{\mathrm{B}} \le P_i^{\max},\ \forall i\in\mathcal{G}^{\mathrm{CC}}\cup\mathcal{G}^{\mathrm{NCC}}\,, \tag{5}$$

$$0 \le w_i \le W_i^{\mathrm{f}},\ \forall i\in\mathcal{G}^{\mathrm{VRG}} \tag{6}$$

$$0 \le l_j \le u_j \le W_j^{\max},\ \forall j\in\mathcal{G}^{\mathrm{VRG}} \tag{7}$$

$$\forall 0 \le \tilde{\boldsymbol{v}} \le 1:\ \exists p_i^{\mathrm{C}}(\tilde{\boldsymbol{v}}), \forall i\in\mathcal{G}^{\mathrm{CC}}:\ s.t. \tag{8}$$

$$\tilde{w}_j = l_j + (u_j - l_j)\cdot\tilde{v}_j,\ \forall j\in\mathcal{G}^{\mathrm{VRG}}\,, \tag{9}$$

$$\sum_{i\in\mathcal{G}^{CC}} p_i^{\mathrm{C}}(\tilde{\boldsymbol{v}}) + \sum_{i\in\mathcal{G}^{\mathrm{NCC}}} p_i^{\mathrm{B}} + \sum_{j\in\mathcal{G}^{\mathrm{VRG}}} \tilde{w}_j = \sum_{n\in\mathcal{I}^{\mathrm{bus}}} D_n \tag{10}$$

$$\left|\sum_{n\in\mathcal{I}^{\mathrm{bus}}} SF_{l,n}\cdot\left(\sum_{i\in\mathcal{G}_n^{\mathrm{CC}}} p_i^{\mathrm{C}}(\tilde{\boldsymbol{v}}) + \sum_{i\in\mathcal{G}_n^{\mathrm{NCC}}} p_i^{\mathrm{B}} + \sum_{j\in\mathcal{G}_n^{\mathrm{VRG}}} \tilde{w}_j - D_n\right)\right| \le F_l^{\max},\ \forall l\in\mathcal{I}^{\mathrm{line}}\,, \tag{11}$$

$$-\Delta_i \le p_i^{\mathrm{C}}(\tilde{\boldsymbol{v}}) - p_i^{\mathrm{B}} \le \Delta_i,\ \forall i\in\mathcal{G}^{\mathrm{CC}}\,, \tag{12}$$

$$P_i^{\min} \le p_i(\tilde{\boldsymbol{v}}) \le P_i^{\max},\ \forall i\in\mathcal{G}^{\mathrm{CC}} \tag{13}$$

Decision variables include the lower ($l_j$) and upper ($u_j$) limits of VRG outputs, the base-case generation output of VRG units ($w_j$) and conventional units ($p_i^{\mathrm{B}}$). The objective to be maximized in (1) represents the probability that the realized available VRG output is within the DNE limits.

Equations (2)-(7) define the operation constraints in the base case. Equation (2) denotes the power balance constraint in the base case. Equation (3) models the transmission capacity constraint. Equation (4) denotes the ramping capability of conventional units from the current period to the next one. Equations (5) and (6) represent the output limits of conventional units and VRG units, respectively. Equation (7) indicates that the upper limits of VRG outputs should not exceed the generation capacity of VRG units and the lower limits are nonnegative.

Equations (8)-(13) require that for any realization of VRG output within the DNE limits, there exists feasible corrective dispatch, denoted by $p_i^{\mathrm{C}}(\tilde{\boldsymbol{v}})$, with which the realized VRG output can be accommodated by the system without compromising reliability. Equation (9) denotes the realized VRG output in as a convex combination of its DNE limits. Equations (10)-(13) define the operation constraints that must be satisfied by the corrective dispatch, including the power balance constraints, transmission capacity constraints, adjustment constraints, and output limit constraints of CCUs in the corrective dispatch, respectively.

*2) OBP optimization problem*. The OBP is to determine an optimal operating point given possible wind power scenarios. There are two perspectives regarding handling the uncertainties. First, we would like to ensure that the operating points are robust for any possible wind realization to ensure a high level of security, so we model this perspective as a robust optimization constraint. The uncertainty information used here is value ranges. Second, we would like to seek the most cost-effective solution among all robust solutions, given possible wind realizations. Therefore, we set the objective function as minimizing the expectation of the production costs. The uncertainty information used here is the distribution functions. Note that this model is hybrid of robust optimization and expectation minimization. Various hybrid models can be found in [10], [12]. The optimal DNE limits, denoted as [$\boldsymbol{l}$*,$\boldsymbol{u}$*], obtained by solving the DNE limit problem, may correspond to multiple feasible OBPs ($\boldsymbol{p}^{\mathrm{B}}$). With the DNE limits fixed to [$\boldsymbol{l}$*,$\boldsymbol{u}$*], the following problem is solved to yield the optimal OBPs ($\boldsymbol{p}^{\mathrm{B}*}$) that minimize the expected total dispatch cost without loss of the robustness within the DNE limits:

**(OBP-1)**
$$\min_{\boldsymbol{w},\boldsymbol{p}^{\mathrm{B}}} \mathbb{E}_{\tilde{\boldsymbol{w}}}\left[Q(\boldsymbol{p}^{\mathrm{B}},\tilde{\boldsymbol{w}})\right], \text{ subject to: } \boldsymbol{l}=\boldsymbol{l}^*,\ \boldsymbol{u}=\boldsymbol{u}^*, \text{ constraints (8)-(13).} \tag{14}$$

Note that the DNE limits [$\boldsymbol{l}$*,$\boldsymbol{u}$*] are input parameters in this problem. Again, constraints (8)-(13) require that the OBPs can still ensure the feasibility of corrective dispatch for any realization of VRG output within the fixed DNE limits.

$Q(\boldsymbol{p}^{\mathrm{B}}, \tilde{\boldsymbol{w}})$ in (14) denotes the optimal dispatch cost of the second-stage corrective dispatch problem:

$$Q(\boldsymbol{p}^{\mathrm{B}},\tilde{\boldsymbol{w}})=\min_{\boldsymbol{p}^{\mathrm{C}}}\sum_{i\in\mathcal{G}^{\mathrm{CC}}}C_i(p_i^{\mathrm{C}})+\sum_{i\in\mathcal{G}^{\mathrm{NCC}}}C_i(p_i^{\mathrm{B}}) \tag{15}$$

$$\text{subject to: } \sum_{i\in\mathcal{G}^{\mathrm{CC}}}p_i^{\mathrm{C}}+\sum_{i\in\mathcal{G}^{\mathrm{NCC}}}p_i^{\mathrm{B}}+\sum_{j\in\mathcal{G}^{\mathrm{VRG}}}\tilde{w}_j=\sum_{n\in\mathcal{I}^{\mathrm{bus}}}D_n, \tag{16}$$

$$\left|\sum_{n\in\mathcal{I}^{\mathrm{bus}}}SF_{l,n}\cdot\left(\sum_{i\in\mathcal{G}_n^{\mathrm{CC}}}p_i^{\mathrm{C}}+\sum_{i\in\mathcal{G}_n^{\mathrm{NCC}}}p_i^{\mathrm{B}}+\sum_{j\in\mathcal{G}_n^{\mathrm{VRG}}}\tilde{w}_j-D_n\right)\right|\le F_l^{\max}\quad\forall l\in\mathcal{I}^{\mathrm{line}}, \tag{17}$$

$$-\Delta_i\le p_i^{\mathrm{C}}-p_i^{\mathrm{B}}\le\Delta_i,\ \forall i\in\mathcal{G}^{\mathrm{CC}}, \tag{18}$$

$$P_i^{\min}\le p_i^{\mathrm{C}}\le P_i^{\max},\ \forall i\in\mathcal{G}^{\mathrm{CC}}. \tag{19}$$

Equations (16)-(19) define the constraints that must be satisfied by the corrective dispatch, including the power balance constraints, transmission capacity constraints, adjustment constraints, and output limit constraints of CCUs in the corrective dispatch, respectively.

### *3.2 Data-Driven Approximation*

Evaluation of the probability in (1) and expectation in (14) requires accurate distribution functions, which are usually not available in practice. Even if the accurate distribution functions are available, the resulting formulation could be too complex to optimize, unless the distribution functions take simple forms. To address this issue, we employ the SAA technique [13] to approximate the probability in (1) and expectation in (14), where the probability distributions are replaced by using historical observations of wind output levels.

*1) Dispatchable range determination problem*. Using SAA, the objective function in (1) is approximately described as:

$$\mathbb{P}\{\boldsymbol{l}\le\tilde{\boldsymbol{w}}\le\boldsymbol{u}\}\approx\frac{1}{|\mathcal{S}|}\sum_{k\in\mathcal{S}}\mathbb{I}(\boldsymbol{l}\le\boldsymbol{W}^{\mathrm{f}}+\boldsymbol{e}^{(k)}\le\boldsymbol{u}) \tag{20}$$

where the vector $\boldsymbol{e}^{(k)}$ denotes the error term of the $k$-th VRG output forecast. The indicator function $\mathbb{I}(\cdot)$ is defined as:

$$\mathbb{I}(\boldsymbol{l}\le\boldsymbol{W}^{\mathrm{f}}+\boldsymbol{e}^{(k)}\le\boldsymbol{u})=\begin{cases}1 & \text{if } l_j\le W_j^{\mathrm{f}}+e_j^{(k)}\le u_j,\ \forall j\in\mathcal{G}^{\mathrm{VRG}}\\ 0 & \text{otherwise}\end{cases} \tag{21}$$

By introducing the binary indicator variables $z^{(k)}$, the **DNE-1** model is approximately reformulated to an MILP as follows:

**(DNE-2)**
$$\min_{\boldsymbol{l},\boldsymbol{u},\boldsymbol{z},\boldsymbol{w},\boldsymbol{p}^{\mathrm{B}}}\sum_{k\in\mathcal{S}^{\mathrm{DNE}}}z^{(k)}$$

subject to: Constraints (2)-(13),

$$(W_j^{\max}-W_j^{\mathrm{f}}-e_j^{(k)})\cdot z^{(k)}-l_j\ge -W_j^{\mathrm{f}}-e_j^{(k)},\ \forall j\in\mathcal{G}^{\mathrm{VRG}},k\in\mathcal{S}^{\mathrm{DNE}}, \tag{22}$$

$$(W_j^{\mathrm{f}}+e_j^{(k)})\cdot z^{(k)}+u_j\ge W_j^{\mathrm{f}}+e_j^{(k)},\ \forall j\in\mathcal{G}^{\mathrm{VRG}},k\in\mathcal{S}^{\mathrm{DNE}}, \tag{23}$$

$$z^{(k)} \in \{0,1\}, \forall k \in \mathcal{S}^{\mathrm{DNE}} \tag{24}$$

From (22) and (23), it can be seen that $l_j \le W_j^{\mathrm{f}} + e_j^{(k)} \le u_j$ if $z^{(k)} = 0$, and $z^{(k)} = 1$ if $W_j^{\mathrm{f}} + e_j^{(k)} > u_j$ or $W_j^{\mathrm{f}} + e_j^{(k)} < l_j$. This indicates that $z^{(k)}$ is equal to zero if the $k$-th data sample is within the DNE ranges and is equal to 1 otherwise. The objective of **DNE-2** is to minimize the number of scenarios outside the DNE limits, or equivalently, to maximize the number of scenarios within the DNE limits. Since the number of scenarios within the DNE limits is an approximation of the probability that the realized available VRG output is within the DNE limits, **DNE-2** is indeed an approximate reformulation of **DNE-1**.

*2) OBP optimization problem.* By introducing the recourse variables $p_i^{\mathrm{C}(k)}$ for the each data sample in $\mathcal{S}^{\mathrm{OBP}}$, the objective function in (14) is approximated by the following:

$$\mathbb{E}_{\tilde{\boldsymbol{w}}}\left[Q(\boldsymbol{p}^{\mathrm{B}}, \tilde{\boldsymbol{w}})\right] \approx \frac{1}{\left|\mathcal{S}^{\mathrm{OBP}}\right|} \sum_{k \in \mathcal{S}^{\mathrm{OBP}}} Q(\boldsymbol{p}^{\mathrm{B}}, \boldsymbol{W}^{\mathrm{f}} + \boldsymbol{e}^{(k)}) \tag{25}$$

where the second-stage corrective dispatch cost $Q(\boldsymbol{p}^{\mathrm{B}}, \boldsymbol{W}^{\mathrm{f}} + \boldsymbol{e}^{(k)})$ is defined in (15)-(19). Hence, the original **OBP-1** model can be approximated by the following:

**(OBP-2)**

$$\min_{\boldsymbol{w}, \boldsymbol{p}^{\mathrm{B}}, \boldsymbol{p}^{\mathrm{C}(\cdot)}} \sum_{i \in \mathcal{G}^{\mathrm{NCC}}} C_i(p_i^{\mathrm{B}}) + \frac{1}{\left|\mathcal{S}^{\mathrm{OBP}}\right|} \sum_{k \in \mathcal{S}^{\mathrm{OBP}}} \sum_{i \in \mathcal{G}^{\mathrm{CC}}} C_i(p_i^{\mathrm{C}(k)})$$

subject to: $\boldsymbol{l} = \boldsymbol{l}^*, \boldsymbol{u} = \boldsymbol{u}^*$, constraints (8)-(13),

$$\sum_{i \in \mathcal{G}^{\mathrm{CC}}} p_i^{\mathrm{C}(k)} + \sum_{i \in \mathcal{G}^{\mathrm{NCC}}} p_i^{\mathrm{B}} + \sum_{j \in \mathcal{G}^{\mathrm{VRG}}} \left(W_j^{\mathrm{f}} + e_j^{(k)}\right) = \sum_{n \in \mathcal{I}^{\mathrm{bus}}} D_n, \forall k \in \mathcal{S}^{\mathrm{OBP}}, \tag{26}$$

$$\left| \sum_{n \in \mathcal{I}^{\mathrm{bus}}} SF_{l,n} \cdot \left( \sum_{i \in \mathcal{G}_n^{\mathrm{CC}}} p_i^{\mathrm{C}(k)} + \sum_{i \in \mathcal{G}_n^{\mathrm{NCC}}} p_i^{\mathrm{B}} + \sum_{j \in \mathcal{G}_n^{\mathrm{VRG}}} \left(W_j^{\mathrm{f}} + e_j^{(k)}\right) - D_n \right) \right| \le F_l^{\max} \quad \forall l \in \mathcal{I}^{\mathrm{line}}, k \in \mathcal{S}^{\mathrm{OBP}}, \tag{27}$$

,

$$-\Delta_i \le p_i^{\mathrm{C}(k)} - p_i^{\mathrm{B}} \le \Delta_i, \forall i \in \mathcal{G}^{\mathrm{CC}}, k \in \mathcal{S}^{\mathrm{OBP}}, \tag{28}$$

$$P_i^{\min} \le p_i^{\mathrm{C}(k)} \le P_i^{\max}, \forall i \in \mathcal{G}^{\mathrm{CC}}, k \in \mathcal{S}^{\mathrm{OBP}}. \tag{29}$$

where (26)-(29) represent the corrective dispatch constraints for all data samples.

Although the SAA technique is employed here, it is different from the conventional one. In the proposed approach, the SAA technique is used for both probability evaluation and expectation calculation. No assumption is made about the probability distribution of VRG forecast errors, and the samples used are real historical data of VRG forecast errors. Clearly, the dataset used in the approximation must be representative and relevant to the situations in the period for which we are scheduling. The data selection strategy is discussed in Section 4.

### *3.3 Strong Extended Formulation*

To improve the solution efficiency, a more compact and tighter formulation of the dispatchable range determination model is developed by using a strong extended formulation technique [14]. The linear relaxation of the resulting reformulation is as strong as having all the valid inequalities for the original formulation [14], [15]. Details of this reformulation technique can be referred in [14].

First, we obtain an estimate of the upper bound of the number of scenarios in $\mathcal{S}^{\mathrm{DNE}}$ that lie outside the DNE ranges, which is denoted by $K$ ( $0 \le K \le |\mathcal{S}^{\mathrm{DNE}}|$ ). To build the strong extended formulation, the sequences $\{W_j^{\mathrm{f}} + e_j^{(k)} : k \in \mathcal{S}^{\mathrm{DNE}}\}$ and $\{W_j^{\max} - W_j^{\mathrm{f}} - e_j^{(k)} : k \in \mathcal{S}^{\mathrm{DNE}}\}$ are sorted in descending order, and the resulting index sequences are denoted by $\{\gamma_1, \gamma_2, ..., \gamma_{|\mathcal{S}^{\mathrm{DNE}}|}\}$ and $\{\phi_1, \phi_2, ..., \phi_{|\mathcal{S}^{\mathrm{DNE}}|}\}$, respectively. The binary variables $\alpha_j^{(k)}$ and $\beta_j^{(k)}$, $k = 1,2,...,K$, $j \in \mathcal{G}^{\mathrm{VRG}}$ are introduced to build the following extended formulation:

**(DNE-3)**

$$\min_{\boldsymbol{l}, \boldsymbol{u}, \boldsymbol{z}, \boldsymbol{w}, \boldsymbol{p}^{\mathrm{B}}} \sum_{k \in \mathcal{S}^{\mathrm{DNE}}} z^{(k)}$$

subject to: Constraints (2)-(13),

$$\sum_{m=1}^{K} \left(e_j^{(\phi_{m+1})} - e_j^{(\phi_m)}\right) \cdot \alpha_j^{(m)} - l_j \ge -W_j^{\mathrm{f}} - e_j^{(\phi_1)}, \forall j \in \mathcal{G}^{\mathrm{VRG}}, \tag{30}$$

$$\alpha_j^{(m)} - \alpha_j^{(m+1)} \ge 0, \forall m = 1,2,...,K-1, j \in \mathcal{G}^{\mathrm{VRG}}, \tag{31}$$

$$z^{(\phi_m)} - \alpha_j^{(m)} \ge 0, \forall m = 1,2,...,K, j \in \mathcal{G}^{\mathrm{VRG}}, \tag{32}$$

$$\alpha_j^{(m)} \in \{0,1\}, \forall m = 1,2,...,K, j \in \mathcal{G}^{\mathrm{VRG}}, \tag{33}$$

$$\sum_{m=1}^{K} \left(e_j^{(\gamma_m)} - e_j^{(\gamma_{m+1})}\right) \cdot \beta_j^{(m)} + u_j \ge W_j^{\mathrm{f}} + e_j^{(\gamma_1)}, \forall j \in \mathcal{G}^{\mathrm{VRG}}, \tag{34}$$

$$\beta_j^{(m)} - \beta_j^{(m+1)} \ge 0, \forall m = 1,2,...,K-1, j \in \mathcal{G}^{\mathrm{VRG}}, \tag{35}$$

$$z^{(\gamma_m)} - \beta_j^{(m)} \ge 0, \forall m = 1,2,...,K, j \in \mathcal{G}^{\mathrm{VRG}}, \tag{36}$$

$$\beta_j^{(m)} \in \{0,1\}, \forall m = 1,2,...,K, j \in \mathcal{G}^{\text{VRG}}, \tag{37}$$

Constraints (30)-(37) are added to the **DNE-3** model to replace the original constraints (22)-(23) in the **DNE-2** model.

Note that if the value of $K$ is underestimated, the **DNE-3** model is infeasible. In this case, we can increase the value of $K$ and solve the problem again.

*3.4 Solution Algorithm for Two-Stage RO Problems*

The **DNE-3** and **OBP-2** models can be regarded as risk-constrained two-stage RO problems [12] with the following abstract form:

$$\min_{x} \boldsymbol{c}^T \boldsymbol{x}$$

$$\text{subject to:} \quad \boldsymbol{A}\boldsymbol{x} \le \boldsymbol{b}, \tag{38}$$

$$\forall \tilde{\boldsymbol{v}} \in [0,1]^n : \exists y(\tilde{\boldsymbol{v}}), s.t. (\boldsymbol{B}_i + \tilde{\boldsymbol{v}}^T \boldsymbol{C}_i)\boldsymbol{x} + \boldsymbol{D}_i \boldsymbol{y}(\tilde{\boldsymbol{v}}) \le e_i, \forall i. \tag{39}$$

Here $\boldsymbol{x}$ represents the first-stage decision variables such as $\boldsymbol{l}$, $\boldsymbol{u}$, $\boldsymbol{w}$, $\boldsymbol{p}^{\text{B}}$, and $\boldsymbol{z}$ for **DNE-3** and $\boldsymbol{w}$, $\boldsymbol{p}^{\text{B}}$, and $\boldsymbol{p}^{\text{C}(.)}$ for **OBP-2**. $\tilde{\boldsymbol{v}}$ represents the uncertain parameters; $\boldsymbol{y}(\tilde{\boldsymbol{v}})$ stands for the second-stage recourse variables (i.e., the corrective dispatch $\boldsymbol{p}^{\text{C}}$ in the robust constraints of both formulations). Equation (39) represents constraints (8)-(13), and (38) denotes other constraints in the models. Note that (39) consists of infinitely many scenarios; hence, it is not possible to explicitly include all scenarios in the formulation. In this study, we propose to add the violated scenarios only when necessary, which is normally called the column-and-constraint generation (C&CG) method [16]. The C&CG algorithm employs a "master-problem and subproblem" iterative framework, in which the master is a relaxation of the original problem with finitely many constraints, and the subproblem identifies the most violated scenario. The violated scenario is sent to the master problem, and the master problem is solved again. The process is repeated until no additional violated scenarios are identified.

For presentation convenience, the sub-problem is defined as follows:

**(SP)** $\quad \Theta(\boldsymbol{x}) = \max_{0 \le \tilde{\boldsymbol{v}} \le 1} \min_{\boldsymbol{y},\boldsymbol{s}} \boldsymbol{1}^T \boldsymbol{s}$, subject to: $\boldsymbol{s} \ge 0$, $\boldsymbol{x}^T \boldsymbol{C}_i \tilde{\boldsymbol{v}}^T + \boldsymbol{D}_i \boldsymbol{y}(\tilde{\boldsymbol{v}}) \le e_i - \boldsymbol{B}_i \boldsymbol{x}, \forall i$

where $\boldsymbol{s}$ denotes a vector of slack variables. If $\Theta(\boldsymbol{x}) = 0$, there are always feasible recourse decisions in response to the realization of random parameters in the uncertainty set. If $\Theta(\boldsymbol{x}) > 0$, there is at least one realization within the uncertainty set for which recourse actions are not feasible.

Because of its special structure, the max-min **SP** problem can be transformed into an equivalent MILP by taking dual formulation of the inner-level linear program and by the reformulation technique [17], and the resultant reformulation model can be solved directly by off-the-shelf solvers. The C&CG procedure of solving the two RO models is presented briefly as follows:

**Step 1**: Initialization. Set the iteration count at $n = 0$ and the convergence threshold at $\varepsilon > 0$.

**Step 2**: Solve the master problem (**MP**) and obtain the optimal solution $\boldsymbol{x}^{*(n+1)}$:

**(MP)** $\quad \min_{\boldsymbol{x},\boldsymbol{y}^{(\cdot)}} \boldsymbol{c}^T \boldsymbol{x}$, subject to: $\boldsymbol{A}\boldsymbol{x} \le \boldsymbol{b}$, $(\boldsymbol{B}_i + \tilde{\boldsymbol{v}}^{*(k)T} \boldsymbol{C}_i)\boldsymbol{x} + \boldsymbol{D}_i \boldsymbol{y}^{(k)} \le e_i, \forall i; k = 1,2,...,n$

**Step 3**: Solve the subproblem (**SP**) with given $\boldsymbol{x} = \boldsymbol{x}^*$, and obtain the optimal solution $\tilde{\boldsymbol{v}}^{*(n+1)}$ and optimal value $\Theta(\boldsymbol{x}^{*(n+1)})$.

**Step 4**: Check convergence. If $\Theta(\boldsymbol{x}^{*(n+1)}) < \varepsilon$, take $\boldsymbol{x}^{*(n+1)}$ as the optimal solution and terminate; otherwise, go to Step 5.

**Step 5**: Generate columns and constraints. Include additional decision variables $\boldsymbol{y}^{(k+1)}$ and the following constraints in **MP**. Set $n = n + 1$ and go to Step 2.

$$(\boldsymbol{B}_i + \tilde{\boldsymbol{v}}^{*(n+1)T} \boldsymbol{C}_i)\boldsymbol{x} + \boldsymbol{D}_i \boldsymbol{y}^{(n+1)} \le e_i, \forall i.$$

## 4. Data Sample Selection Strategy

The data sample sets $\{\boldsymbol{e}^{(k)} : k \in \mathcal{S}^{\text{DNE}}\}$ and $\{\boldsymbol{e}^{(k)} : k \in \mathcal{S}^{\text{OBP}}\}$ are critical input parameters of the proposed models. Since the data samples are not involved in constraints (8)-(13), they have no impact on the reliability guarantee within the DNE limits. However, as shown in the formulation of the **DNE-3** and **OBP-2** models, the data samples do affect the quality of final solutions. For the dispatchable range determination problem, data samples affect the probability that the DNE ranges cover the actual available VRG output. For the OBP optimization problem, data samples have an influence on the economic efficiency of the OBPs. In order to obtain high-quality solutions, the data sample sets need selecting carefully. This section focuses on the data sample selection strategy.

Assume that a sequence of raw historical data on forecast values and prediction errors of VRG output, denoted by the set $\{(\boldsymbol{W}^{\text{f}(k)}, \boldsymbol{e}^{(k)}) : k \in \mathcal{D}^{\text{hist}}\}$, is accessible to the system operators and that the VRG forecast values and errors in the upcoming period are $(\boldsymbol{W}^{\text{f}}, \boldsymbol{e})$. Since the distribution of VRG forecast errors is recognized to be conditional to the level of forecast values [18]-[19], it is reasonable to believe that the closer the historical forecast value $\boldsymbol{W}^{\text{f}(k)}$ is to the future forecast $\boldsymbol{W}^{\text{f}}$, the more likely the historical forecast error $\boldsymbol{e}^{(k)}$ is to follow the distribution of $\boldsymbol{e}$. Based on that, the following strategy is developed to select a predefined number ($N^{\text{S}}$) of samples from the historical data:

**Step 1**: Calculate the Euclidean distances between each of the historical forecast values and the upcoming forecast values:

$$\text{dist}(\boldsymbol{W}^{\text{f}(k)}, \boldsymbol{W}^{\text{f}}) = \sqrt{\sum_{j \in \mathcal{G}^{\text{VRG}}} (W_j^{\text{f}(k)} - W_j^{\text{f}})^2}, \forall k \in \mathcal{D}^{\text{hist}}$$

**Step 2**: Sort the sequence $\{\text{dist}(\boldsymbol{W}^{f(k)},\boldsymbol{W}^{f}):k\in\mathcal{D}^{\text{hist}}\}$ in ascending order and obtain the index sequence $\{\nu_1,\nu_2,...,\nu_{|\mathcal{D}^{\text{hist}}|}\}$.

**Step 3**: Select the first $N^{s\prime}=\min\{N^{s},|\mathcal{D}^{\text{hist}}|\}$ elements in the sorted index sequence to form the data sample set $\mathcal{S}$:

$$\mathcal{S}=\{\nu_k:k=1,2,...,N^{s\prime}\}$$

## 5. Numerical Tests

Numerical tests are conducted with two test systems to study the performance of the proposed models and approaches. The first case study is done by using a modified 6-bus system with 2 wind farms to show how the proposed models affect the economic benefits and wind power integration. The second is carried out by using a modified 118-bus system with 10 wind farms to verify the scalability of the proposed approaches. Test system data are described in details in [20].

Wind power data, including predictions and observations, are extracted from the Eastern Wind Dataset provided by the National Renewable Energy Laboratory (NREL) [21]. The dataset includes hourly wind data on more than 1,000 wind farms in the Eastern U.S. from 2004 to 2006. All simulations are programmed on a personal computer running at 2.40 GHz with 16 GB of memory. The simulation platform is Matlab R2013a, and the Gurobi solvers [22] are launched to solve LPs and MILPs.

The proposed dispatch methodology is compared with the original DNE-based approach in [11], which maximizes the sum of ranges weighted by LMPs:

(**ODNE**) $$\min_{\boldsymbol{l},\boldsymbol{u}} \sum_{j\in\mathcal{G}^{\text{VRG}}} LMP_j\cdot(u_j-l_j)$$, subject to: Constraints (7)-(13).

### *5.1 6-Bus Test System*

As shown in Figure 2, the 6-bus test system consists of 6 buses, 7 lines, 2 conventional units (G1 and G2), and 2 wind farms (W1 and W2). Wind data for two wind farms, including forecast values and observed output, are obtained from data on Sites 3902 and 3945 in the NREL Eastern Wind Dataset, which are scaled by a factor of 0.1.

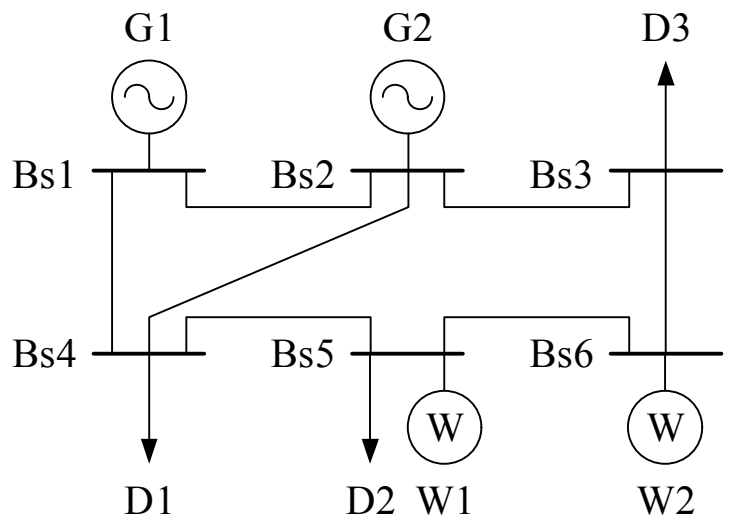

Figure 2. 6-bus test system.

The proposed real-time dispatch approach is compared with the **ODNE** method using dispatch simulation. The simulation procedures for the two approaches are described as follows:

*1) The proposed method.* Hourly dispatch is simulated in a receding horizon. At the beginning of each period, DNE limits of VRG units are first computed by solving the **DNE-3** model with forecast information for the next period. Given the DNE limits, OBPs of conventional units are then calculated by solving the **OBP-2** model. After the wind farm outputs are observed (set to their real observations), corrective dispatch is performed, and benefits are evaluated. Then the time period is moved a step forward, and the hour dispatch is performed again. In this simulation, hourly wind data from January 1 through December 31, 2004, including up to 8,783 data points, are employed as the raw historical dataset. Wind data from January 1 to 10, including 240 data points, are used as the validation data for dispatch simulation. In each run of simulation, $N^{\text{DNE}}$ = 400 samples and $N^{\text{OBP}}$ = 20 samples are selected from the raw historical dataset to form the data sample sets $\mathcal{S}^{\text{DNE}}$ and $\mathcal{S}^{\text{OBP}}$, respectively.

*2) ODNE method.* Hourly dispatch is conducted in a receding horizon. At each period, OBPs of all units are first calculated by solving a conventional ED model using forecast information for the next period. Given the OBPs, DNE limits of wind farms are then determined by solving the **ODNE** model. After the wind farm outputs are observed, corrective dispatch is performed. The scheduling period is then moved a step forward to start the next turn of power dispatch.

The DNE limits of each wind farm obtained by two different approaches are displayed in Figure 3. For the DNE method, the LMP at Bus 6 is higher than that at Bus 5 most of the time. As a result, wider DNE ranges are allocated to W2 in order to maximize the sum of weighted DNE ranges, and the DNE ranges of W2 are usually wider than those of W1. Consequently, the observed available wind power output cannot be fully captured by the ODNE ranges at many periods. Only 57.5% of the available wind power outputs are captured by the ODNE ranges. In contrast, the proposed method allows the DNE ranges to be allocated more evenly, and up to 95.4% of the wind power availability realizations are within the resultant DNE limits.

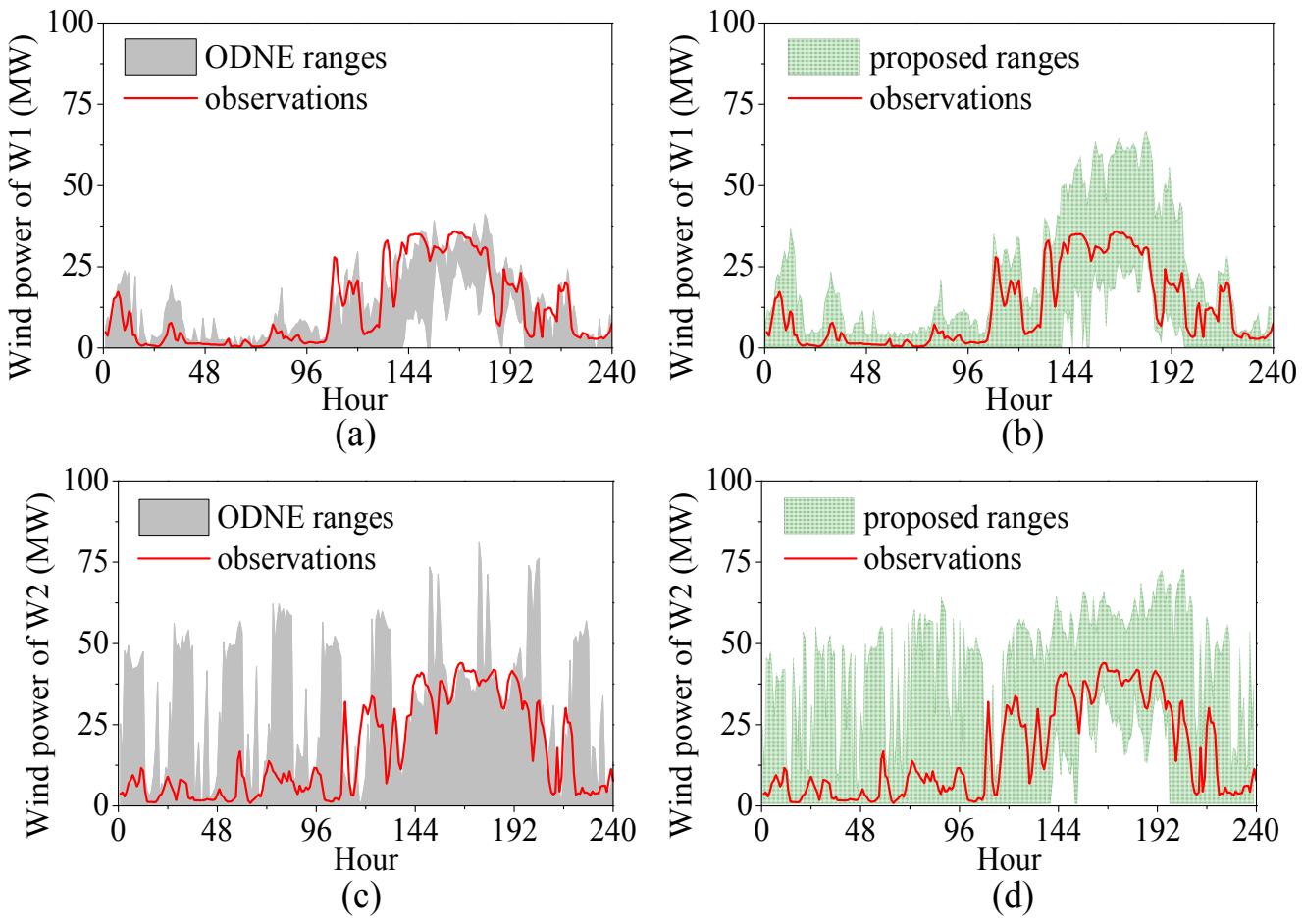

Figure 3. DNE limits and observed wind power outputs of wind farms by the ODNE model (a and c) and the proposed model (b and d).

Figure 4 shows the DNE limits obtained by both approaches as well as data samples at $t = 168$, where the cross markers denote the selected data samples at that snapshot. It can be seen that the DNE ranges by the proposed method (red box) cover most of the samples, while the ODNE ranges (blue box) cover only a small portion of the samples. The enhanced rate of coverage by the DNE ranges is due to the use of data samples and the co-optimization between DNE limits and OBPs. As the **DNE-3** model aims to minimize the number of points that lie outside the DNE range, the resultant DNE limits are able to contour more samples than the ODNE limits can. Since the data samples represent the true probability distribution of available wind power only approximately, the actual wind power output is more likely to be captured by the DNE ranges of the proposed method than by those of the ODNE method. In addition, it can be seen that the red box is larger than the blue one, which indicates that the DNE range obtained by the proposed method allows more flexibility than the range obtained by the other method. This is because in the **DNE-3** model, DNE limits are co-optimized with variable OBPs, while the OBPs are fixed in the **ODNE** model. As a result, with the wind power availability shown in the figure, the wind power of both wind farms has to be unnecessarily curtailed by the ODNE method, but this does not happen when the proposed method is used.

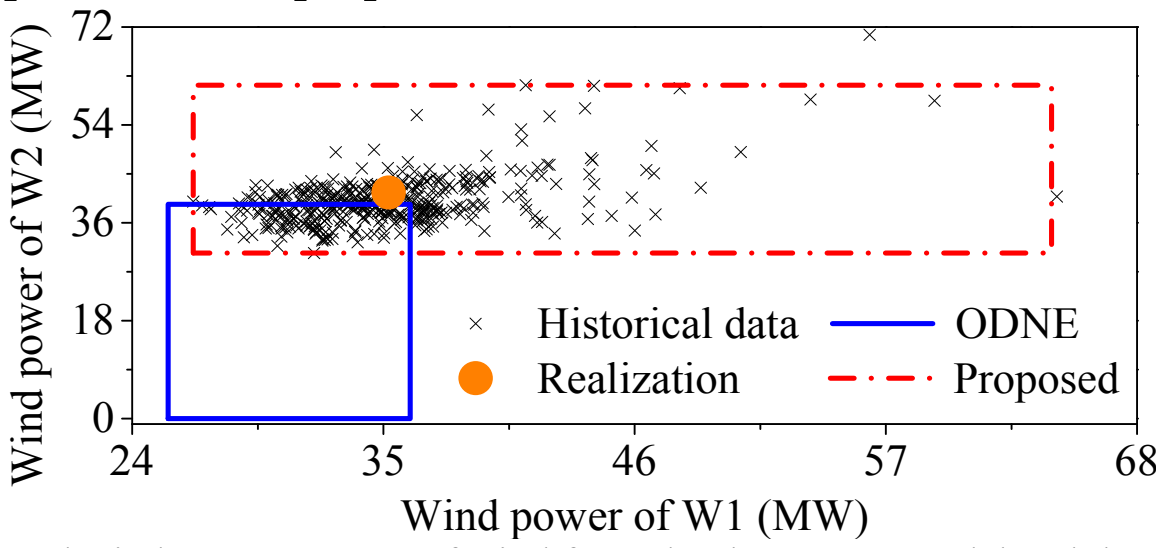

Figure 4. Data samples, DNE limits, and observed wind power outputs of wind farms by the ODNE model and the proposed method at snapshot $t = 168$.

The total wind power output by two methods in each period is shown in Figure 5(a). The wind power accommodated by the proposed approach is more than that by the ODNE method in nearly all the periods, and the difference in wind power output is sometimes large. This result indicates that the proposed method is able to promote better wind power utilization.

Figure 5(b) shows the total dispatch costs of two methods in all periods. In most periods, the proposed method requires less total dispatch cost than the other one. In other periods, dispatch by the proposed method is more expensive than dispatch by the ODNE method. In the **DNE-3** model of the proposed approach, DNE limits and OBPs are co-optimized to maximize the coverage rate, which represents the potential ranges of wind power generation. In order to accommodate more wind power, conventional units need to spare more regulation reserve capacity to cope with potential wind power volatility. As a result, conventional units would have to be operated in a less economic manner, which is the price of gaining more flexibility. Therefore, the dispatch cost of the proposed method could sometimes be higher than that of the ODNE method, though wind power generation is promoted by the proposed method. This result also indicates that promoting VRG is not always a cost-effective choice.

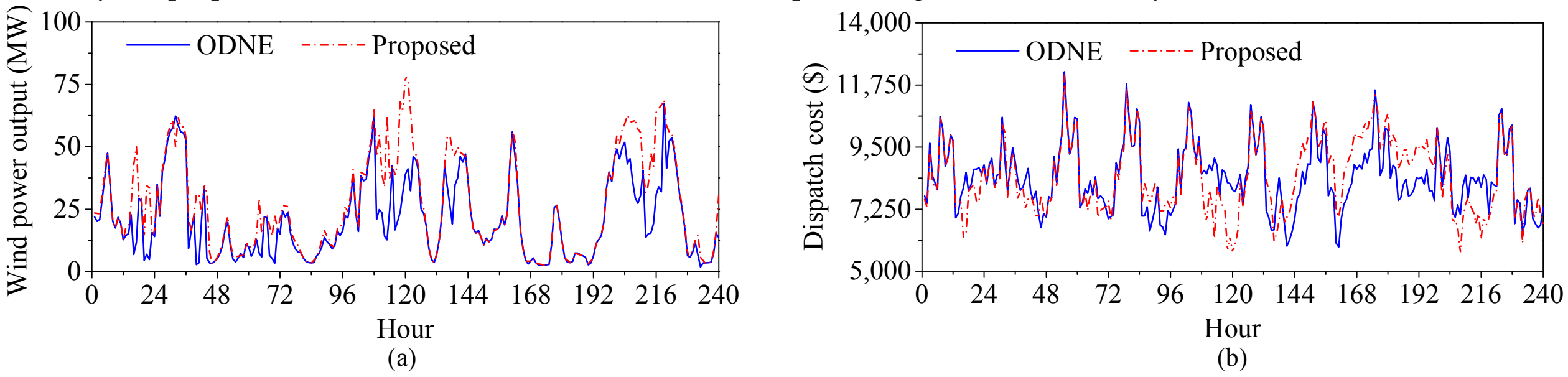

Figure 5. (a) Total wind power output and (b) total costs of dispatch simulations by two approaches.

### 5.2 118-Bus Test System

The modified 118-bus test system consists of 186 lines, 76 conventional units, and 10 wind farms. Wind data for this test system are also extracted from 10 sites in the NREL Eastern Wind Dataset. By using this test system, dispatch simulations are conducted to find the performance differences between the proposed method and the ODNE method. Figure 6(a) shows the total wind power output in MW in each period. Figure 6(b) shows the relative values of total dispatch costs for two methods, where the results of the ODNE method are referenced. It can be seen that the total wind power generation by the proposed method is no less than by the other approach for all periods, and the utilization of wind power is enhanced by the proposed method in most periods. Due to the improved wind power integration, the total dispatch costs are also reduced by the proposed method in most periods, as compared with the case by the ODNE method. This example again reveals the better performance of the proposed method over the other one and indicates the scalability of the proposed method.

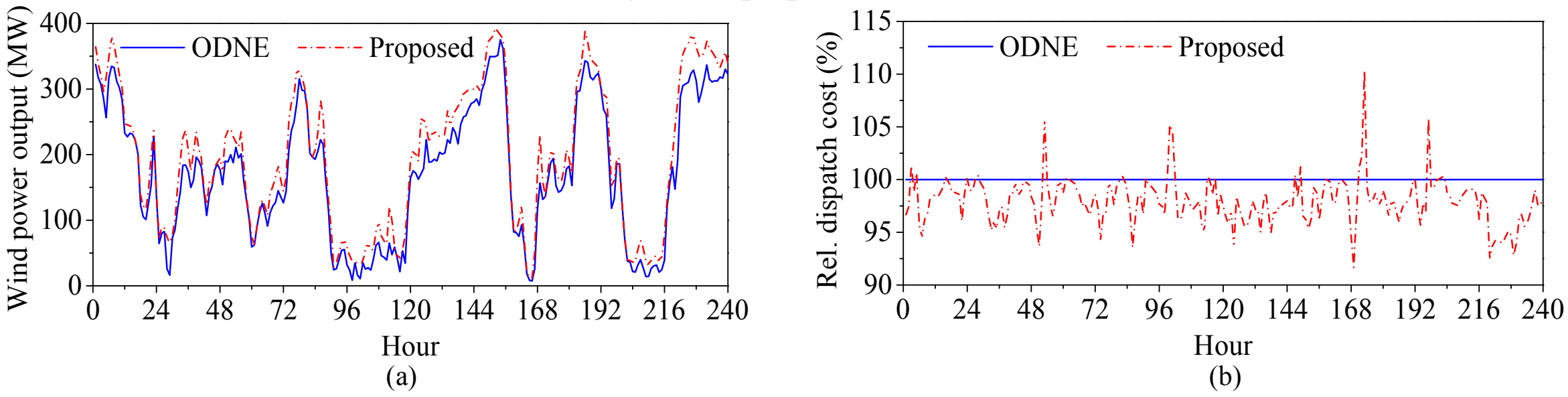

Figure 6. (a) Total wind power output and (b) relative total dispatch costs by two approaches (118-bus).

The impact of the numbers of data samples is studied using this system. For notational simplicity, the numbers of data samples in $\mathcal{S}^{\mathrm{DNE}}$ and $\mathcal{S}^{\mathrm{OBP}}$ are denoted by $N^{\mathrm{DNE}}$ and $N^{\mathrm{OBP}}$, respectively. First, the dispatch simulation is conducted with varying $N^{\mathrm{DNE}}$ and fixed $N^{\mathrm{OBP}} = 20$ using the proposed method. Relative values of hourly total wind power output and total dispatch cost of each case are shown in Figure 7(a) and Figure 7(b), respectively, where the results in case $N^{\mathrm{DNE}} = 400$ and $N^{\mathrm{OBP}} = 20$ are referenced. The average total dispatch cost, average wind power output and average CPU time versus different values of $N^{\mathrm{DNE}}$ is plotted in Figure 8.

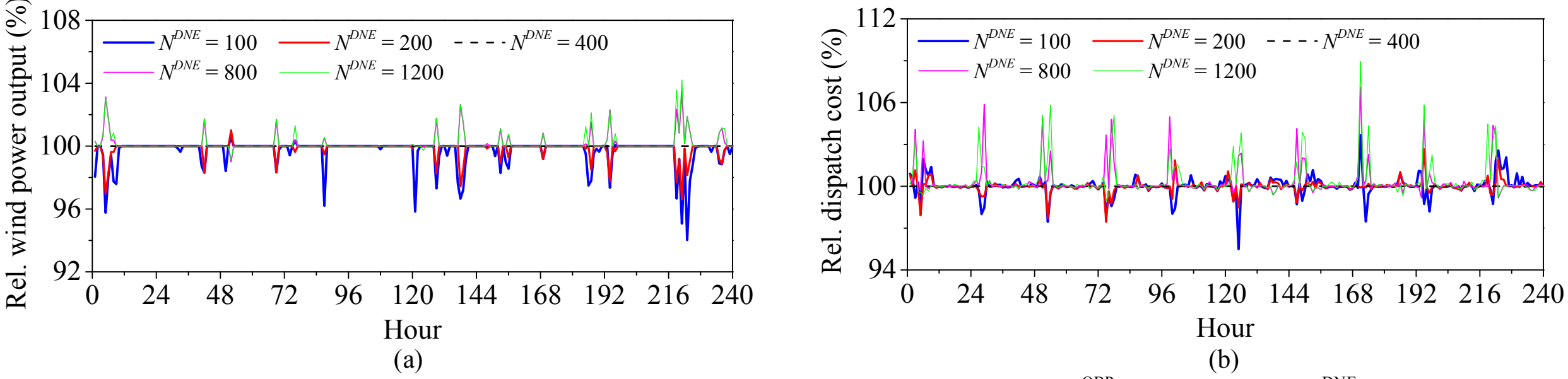

Figure 7. (a) Relative total wind power output and (b) relative total cost by the proposed method with fixed $N^{\mathrm{OBP}} = 20$ and varying $N^{\mathrm{DNE}}$ (118-bus).

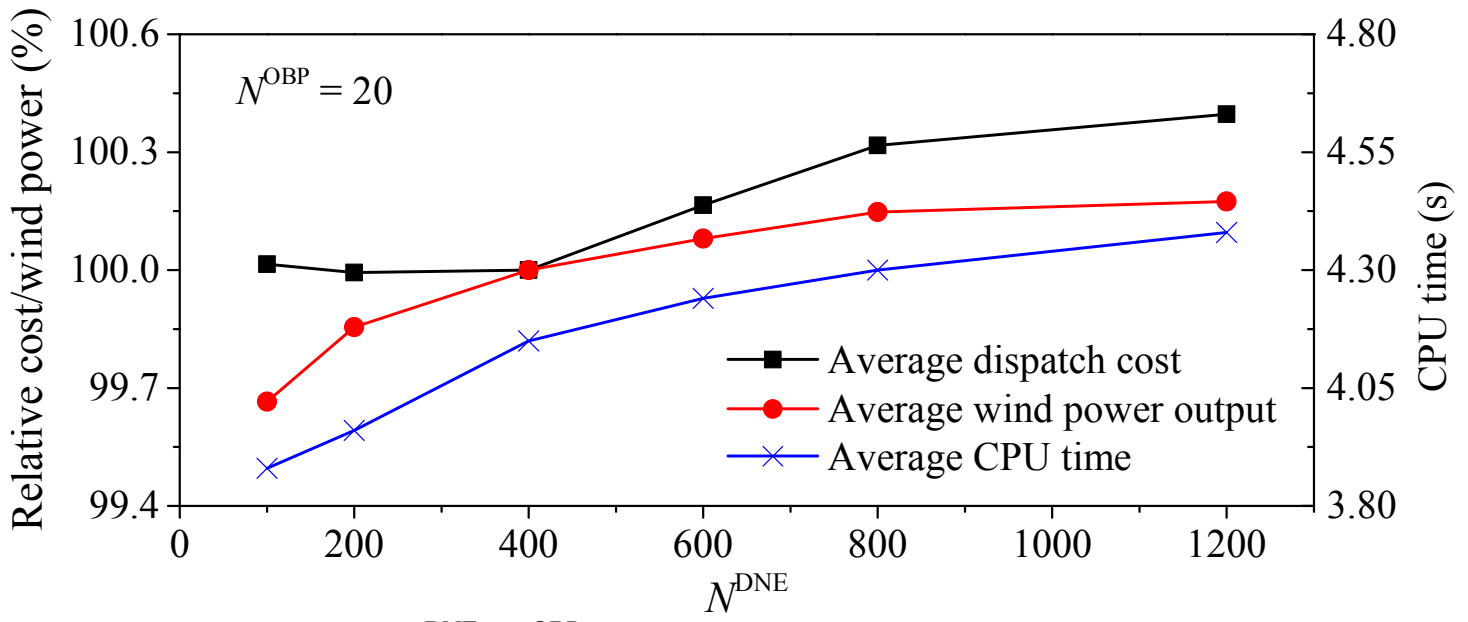

Figure 8. Performance of the proposed method with varying $N^{\mathrm{DNE}}$ ($N^{\mathrm{OBP}} = 20$).

As shown in Figure 7(a), as more data samples are used in the dispatchable range determination, the amount of total wind power output tends to increase in many periods. Figure 8 shows that the average total wind power generation rises up as $N^{\mathrm{DNE}}$ is increased. This is because as more data samples are utilized, the stochastic nature of wind power can be better captured, thus wind power generation can be accommodated more effectively. Meanwhile, it can be seen that the marginal benefits to wind power output brought by increasing $N^{\mathrm{DNE}}$ decreases. In Figure 8, when $N^{\mathrm{DNE}} \geq 800$, increasing $N^{\mathrm{DNE}}$ does not substantially improve the wind power utilization. This is because when the sample set is sufficiently large, most of stochastic nature of wind power is described, and the probability that additional samples introduces new uncertainty information is small.

However, the dispatch costs are not necessarily reduced when more samples are used for DNE limit calculation. As shown in Figure 7(b), a larger value of $N^{\mathrm{DNE}}$ could lead to more expensive dispatch in some periods. This effect could result in a higher average total dispatch cost, as indicated in Figure 8. As more data samples are involved, some extreme wind power scenarios are covered by the DNE ranges. In order to cope with these scenarios, conventional units would have to preserve regulation capacity, which could lead to uneconomic operation of these units.

From Figure 8, it is observed that the **DNE-3** model becomes computationally more expensive when a larger number of data samples are included in $\mathcal{S}^{\text{DNE}}$. This is because the number of constraints and binary variables are linearly proportional to the number of samples in $\mathcal{S}^{\text{DNE}}$. The average CPU time increases with $N^{\text{DNE}}$ in a sublinear manner, as shown in Figure 8.

To summarize, the increase of $N^{\text{DNE}}$ can contribute to better use of VRG but cause more computational burden, and potentially have a negative influence on the total dispatch cost.

The dispatch simulation is performed with varying $N^{\text{OBP}}$ and fixed $N^{\text{DNE}} = 400$ using the proposed method. Figure 9 shows the average dispatch cost, average wind power output and average CPU time versus varying $N^{\text{OBP}}$. Since the data samples in $\mathcal{S}^{\text{OBP}}$ do not affect the DNE limit calculation, the DNE limits obtained in all cases are identical, and the average total wind power generation is also the same in all cases, as shown in Figure 9.

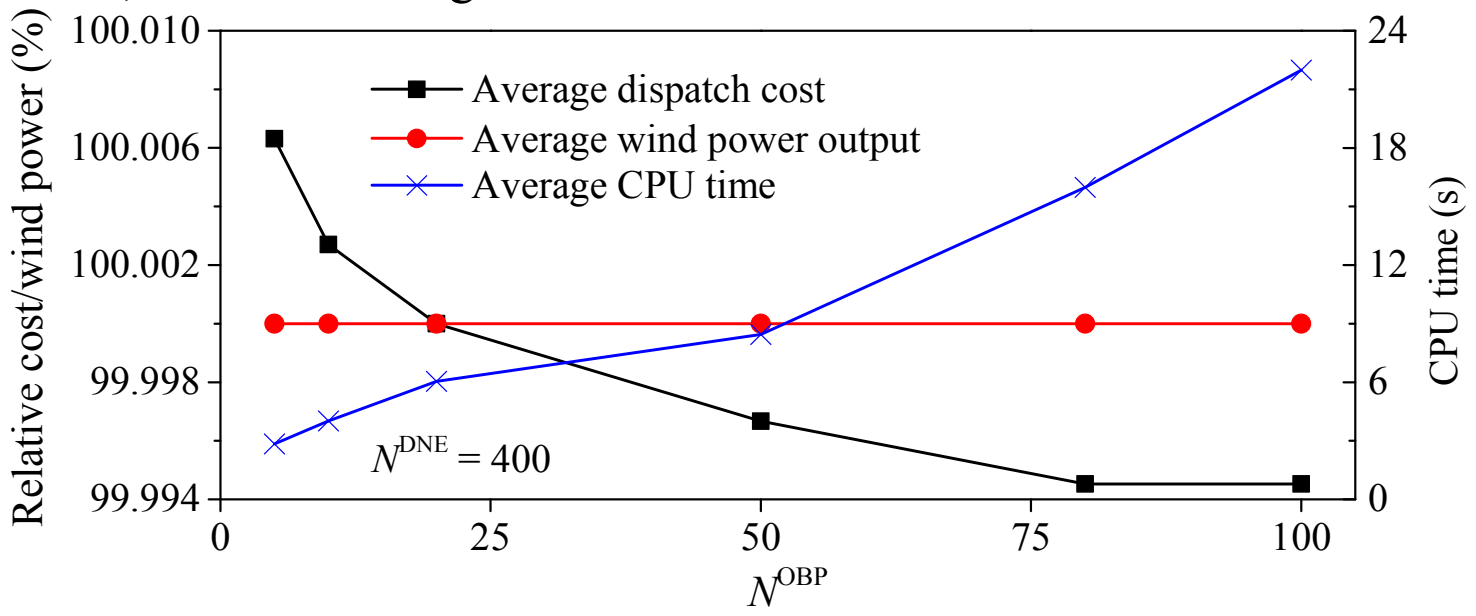

Figure 9. Performance of the proposed method with varying $N^{\text{OBP}}$ ($N^{\text{DNE}} = 400$).

Figure 10 displays the hourly total wind power output with varying $N^{\text{OBP}}$ in terms of relative values, where the results for case $N^{\text{DNE}} = 400$ and $N^{\text{OBP}} = 20$ are referenced. It is shown that as more data samples are included in $\mathcal{S}^{\text{OBP}}$ for OBP optimization, less total dispatch cost is needed in many periods. This is consistent with the results in Figure 9 that the average dispatch cost decreases with increasing $N^{\text{OBP}}$. The decreasing rate of average dispatch cost diminishes as $N^{\text{OBP}}$ increases.

On the other hand, more average CPU time is need for solving the OBP optimization problem when $N^{\text{OBP}}$ is enlarged, as shown in Figure 9. It's observed that the average CPU increases approximately linearly with $N^{\text{OBP}}$. This is because the size of the **OBP-2** model is increased linearly with the number of scenarios, and this model becomes more expensive to solve when more variables and constraints are involved.

Therefore, a tradeoff between solution quality and computational burden should be made when the number of samples in $\mathcal{S}^{\text{OBP}}$ is selected.

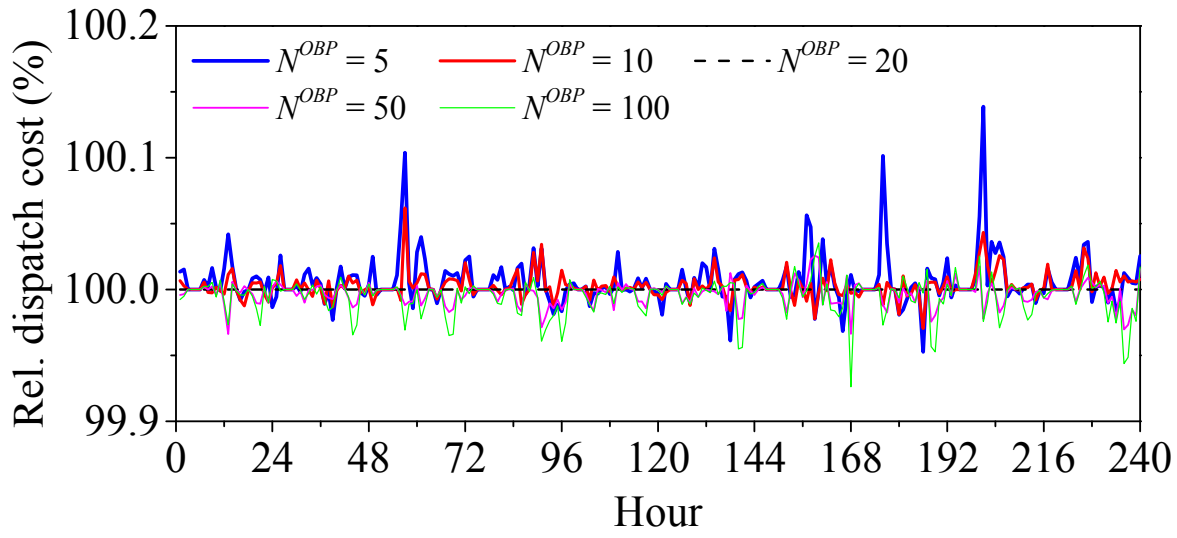

Figure 10. Relative total wind power output by the proposed method with fixed $N^{\text{DNE}} = 400$ and varying $N^{\text{OBP}}$ (118-bus).

## 6. Conclusions

In this paper, a data-driven real-time power dispatch methodology is proposed to maximize the utilization of VRG by using historical VRG forecast error data. A real-time dispatch framework is developed to co-optimize the DNE limits of VRG units and the OBPs of conventional units through solving the dispatchable range determination and the OBP optimization problems. Both problems are formulated by using hybrid stochastic and robust optimization, and their models are approximated by MILPs using the SAA technique based on real historical data samples. Extended formulation is employed to make the MILP model of dispatchable range determination problem more compact and tighter. A data selection strategy for extracting an appropriate dataset from a raw historical data pool is also discussed. Numerical simulations are performed using two test systems, and the test results illustrate that the proposed method can capture the possible wind power available better than the ODNE method does, and that it has great potential for improving wind power integration. It is shown that the proposed method is scalable to large power systems. In addition, the effect of the number of data samples on the proposed method is also discussed and analyzed.

It is noteworthy that the proposed data-driven power dispatch method is not limited to electric power systems. This approach can also be extended to handle power dispatch problems of integrated energy systems [23]-[24] with significant uncertainties pertaining to renewable energy or heat and cooling load demand. Uncertainty management using the data-driven dispatch approach is an interesting topic that is worth further investigation.